\documentstyle[amssymb,aps,multicol,epsf]{revtex}

\begin{document}
\draft
\title{Ising Model on Networks with an Arbitrary Distribution of Connections }
\author{S.N. Dorogovtsev$^{1,2,\ast}$, A.V. Goltsev$^{2,\dagger}$, and J.F.F. Mendes$%
^{1,\ddagger}$}
\address{
$^1$ Departamento de F\'{\i }sica and Centro de F\'{\i }sica do Porto,
Faculdade de Ci\^{e}ncias, Universidade do Porto \\
Rua do Campo Alegre 687, 4169-007 Porto, Portugal\\
$^{2}$ A.F. Ioffe Physico-Technical Institute, 194021 St. Petersburg, Russia}
\maketitle

\begin{abstract}
We find the exact critical temperature $T_c$ of the nearest-neighbor
ferromagnetic Ising model on an `equilibrium' random graph with an arbitrary
degree distribution $P(k)$. We observe an anomalous behavior of the
magnetization, magnetic susceptibility and specific heat, when $P(k)$ is
fat-tailed, or, loosely speaking, when the fourth moment of the distribution
diverges in infinite networks. When the second moment 
becomes divergent, $T_c$ approaches infinity, the phase transition is of
infinite order, and size effect is anomalously strong.
\end{abstract}


\begin{multicols}{2}
\narrowtext


The Ising model is one of the immortal themes in physics. It is a
traditional starting point for the study of the effects of cooperative
behavior. Networks with complex architecture display a spectrum of unique
effects \cite{ba99,s01,ab01a,dm01c,ajb00d,asbs00}, and so they are an
intriguing substrate. The simulation of the Ising model on a growing
scale-free network \cite{ahs01} has demonstrated that it is extremely far
from that on regular lattices and on `planar graphs' \cite{k86}.

In this Letter, we report our exact results and results, which, we believe,
are asymptotically exact, for the thermodynamic properties of the Ising
model on the basic construction for `equilibrium' random networks. 
These networks are the undirected graphs, maximally random under the
constraint that their degree distribution is a given one, $P(k)$. Here, {\em %
degree} is the number of connections of a vertex. Correlations between
degrees of vertices in such graphs are absent. In graph theory, these
networks are called `labelled random graphs with a given degree sequence'
or `the configuration model' \cite{bbk72}. The earlier interest was mainly
in the percolation properties of complex networks and the spread of diseases
on them \cite{mr95,nsw00,ceah00a,cnsw00,pv01,dm01e,cbh02,pv02}, and most of
the analytical results were obtained just for this basic construction
(however, see Refs. \cite{chk01,dms01f,kkkr02a}, where the
Berezinskii-Kosterlitz-Thouless percolation phase transition was studied in
growing networks).

Our results demonstrate the strong effect of the fat tail of the degree
distribution on the phase transition in the Ising model. The most connected
vertices induce strong ferromagnetic correlations in their close
neighborhoods at very large temperatures, and so their role is very
important. Surprisingly, we observe very strong effects of these vertices,
even when they, at first sight, must be insignificant, namely, when the
first and the second moments of the degree distribution are still finite but
the fourth moment already diverges ($\langle k^4 \rangle \to \infty$).

It is convenient to use the power-law degree distribution $P(k) \propto
k^{-\gamma}$ for parametrization. Then, $\langle k^4 \rangle$ diverges for $%
\gamma \leq 5$, $\langle k^3 \rangle$ diverges for $\gamma \leq 4$, and $%
\langle k^2 \rangle$ is divergent for $\gamma \leq 3$. When $\langle k^4
\rangle<\infty$, the phase transition is similar to that in the Ising model
on high-dimension regular lattices. However, for $\gamma<5$ its nature is
quite different. As $\gamma$ decreases and the role of the highly connected
vertices turns to be more important, $T_c$ grows and the phase transition
becomes more `continuous'. Below $\gamma=4$ it 
is of the higher order than the second in Ehrenfest's terminology.
Furthermore, at $\gamma=3$, $T_c$ approaches $\infty$, and the order of the
phase transition is infinite.

When $\langle k^2\rangle $ diverges ($\gamma \leq 3$), the effect of the
highly connected vertices is crucial. In the infinite network, long-range
magnetic order is not destroyed by any temperature, and finite-size effect
is very strong. Formally speaking, we develop a theory for the infinite
networks, but it also allows us to describe finite-size effect. Our main
results are presented in Tab. \ref{t1}.

{\em Intuitive arguments}.---The large networks that are studied in this
paper have a tree-like `local structure'. When we start from a randomly
chosen vertex, and add its first nearest neighbors, second, \ldots, $n$-th
with all their connections, the resulting sub-graph is almost surely a tree.
Of course, in a 
finite size network, there is a boundary for $n$, above which loops appear
in such a sub-graph. 


Cooperative properties, which we study in the present Letter, are determined
by this 
`local' environment, where vertices have a quite different structure of
connections than in the entire net. Interactions are transmitted through
edges, from vertex to vertex. Hence, the following characteristic is
crucial: the distribution of the number of connections of the nearest
neighbor of a vertex. In the networks under consideration, it is $%
kP(k)/\langle k\rangle $. Then, the nearest neighbors of a vertex have the
average number of connections (the average degree) $\langle k^2\rangle
/\langle k\rangle $, its second nearest neighbors have the same average
degree, and so on. Notice that this value is greater than the average number
of connections for the entire network $\langle k\rangle $, and it is much
greater than $\langle k\rangle $ if $\langle k^2\rangle $ is large. 
Therefore, we 
estimate the critical temperature of the Ising model on the network using
the formula $T_c/J=2/\ln [q/(q-2)]$ for the Ising model on a regular Cayley
tree \cite{bbook82} with the coordination number $q=\langle k^2\rangle
/\langle k\rangle $. The result is

\begin{equation}
\frac J{T_c}=\frac 12\ln \left(\frac{\langle k^2\rangle }{\langle k^2\rangle
-2\langle k\rangle }\right)\,.  \label{Tc}
\end{equation}
We will show below that this naive estimate is exact.


\end{multicols} 

\widetext

\begin{center}
\begin{table}[tbp]
\begin{tabular}{|l|l|l|l|l|}
& \ \ \ \ $M$ & \ \ \ \ \ $\delta C(T<T_c)$ & $\!\!\!\!\!\!\!\!\!\chi$ & $%
T_c $ \\ \hline\hline
$\gamma >5,$ $\langle k^4\rangle <\infty $ & $\propto \tau^{1/2}$ $%
\phantom{\frac{B^B}{B^B}}$ & $\!\!\!\!\!\!\!\!\!\!\!\!\!\!\!\!\!\!\!\!\!\!\!%
\!$ jump at $T_c$ decreases as $\langle k^4 \rangle$ grows &  &  \\ 
\cline{1-3}
$\gamma =5,$ $\langle k^4 \rangle =\infty ,$ $\langle k^2 \rangle < \infty $
& $\!\!\!\!\!\!\propto \tau^{1/2}/(\ln \tau^{-1})^{1/2}$ & \ \ $\propto
1/\ln \tau^{-1}$ & $\!\!\!\!\!\!\!\!\!\!\!\!\!\!\propto \tau^{-1}$ & $%
\!\!\!\!\!\!\!\!\!\!\!\!\!\!\!\! 2/\ln\frac{\langle k^2 \rangle} {\langle
k^2 \rangle -2 \langle k \rangle }$ \\ \cline{1-3}
$3<\gamma <5,$ $\langle k^4\rangle =\infty ,$ $\langle k^2\rangle <\infty $
& $\propto \tau^{1/(\gamma -3)}$ $\phantom{\frac{B^{B^{-}}}{B^B}}$ & \ \ $%
\propto \tau^{(5-\gamma)/(\gamma -3)}$ &  &  \\ \hline
$\gamma =3,$ $\langle k^2\rangle =\infty $ & $\propto e^{-2T/\langle k
\rangle}$ $\phantom{\frac{B^{a^{-}}}{B^B}}$ & \ \ $\propto T^2e^{-4T/\langle
k \rangle}$ &  & \ \ $\!\!\!\!\!\!\!\!\!\!\!\!\!\!\!\!\!\!\!\!\!\!\!\!%
\propto \langle k \rangle\ln N$ \\ \cline{1-3}\cline{5-5}
$2<\gamma <3,$ $\langle k^2 \rangle =\infty $ & $\propto T^{-1/(3-\gamma )}$
& \ \ $\propto T^{-(\gamma -1)/(3-\gamma )}$ $\phantom{\frac{B^{B^b}}{B^B}}$
& $\!\!\!\!\!\!\!\!\!\!\!\!\!\!\!\!\! \propto T^{-1}$ & \ \ $\!\!\!\!\!\!\!\!\!\!\!%
\!\!\!\!\!\!\!\!\!\!\!\!\!\propto \langle k \rangle N^{(3-\gamma)/(\gamma
-1)}$%
\end{tabular}
\caption{ Critical behavior of the magnetization $M$, the specific heat $%
\delta C$, and the susceptibility $\chi$ in the Ising model on networks with
a degree distribution $P(k)\sim k^{-\gamma }$ for various values of exponent 
$\gamma $. $\tau \equiv 1-T/T_c$. The right column represents the exact
critical temperature in the case $\langle k^2 \rangle<\infty$ and the
dependence of $T_c$ on the total number $N$ of vertices in a network. }
\label{t1}
\end{table}
\end{center}

\vspace{-40pt}$\phantom{.}$

\begin{multicols}{2}
\narrowtext


{\em General solution}.---Consider the Ising model on an a network with the
Hamiltonian: ${\cal H}=-J\sum_{\langle ij\rangle }S_iS_j-H\sum_iS_i$, where
the first sum is over all edges of the graph, the second one is over all
vertices, $J>0$ and $H$ are the energy of the ferromagnetic interaction and
magnetic field, respectively. Hereafter, we set $J=1$. It is known that the
regular Cayley tree is solved exactly by using recurrence relations \cite
{bbook82}. As networks under discussion have a local tree-like structure, we
apply this method to the Ising model on such nets. Consider spin $S_0$ on a
vertex $0$ with $k_0$ adjacent spins $S_{1,i}$, $i=1,2...k_0$. Due to the
local tree-like structure this spin may be treated as a root of a tree. We
introduce

\begin{equation}
g_{1,i}(S_0)=\!\sum_{S_l=\pm 1}\!\!\!\exp \!\left[ \!\left(
\sum_{\left\langle lm\right\rangle }S_lS_m+S_0S_{1,i}+H\sum_lS_l\!\right)
\!\!/T\right] ,  \label{gg}
\end{equation}
where $T$ is temperature. The indices $l$ and $m$ run only over spins that
belong to sub-trees with the root spin $S_{1,i}$, including $S_{1,i}$. Let $%
x_{1,i}\equiv g_{1,i}(-)/g_{1,i}(+)$, then the magnetic moment $M$ of the
vertex $0$ is 

\begin{equation}
M=(e^{2H/T}-\prod_{i=1}^{k_0}x_{1,i})/(e^{2H/T}+\prod_{i=1}^{k_0}x_{1,i})\,.
\label{M}
\end{equation}
The parameters $x_{1,i}$ describe the effects of the nearest neighbors on
the spin $S_0$. 
In turn, $x_{1,i}$ are expressed in terms of 
parameters $x_{2,l}=g_{2,l}(-)/g_{2,l}(+)$, $l=1,2...k_{1,i}$, which
describe 
effects of spins in the second shell on spins in the first shell, and so on.
The following recurrence relation between $x_{n,j}$ and $x_{n+1,l}$ holds 

\begin{equation}
x_{n,j} = y\left(\prod_{l=1}^{k_{n,j}-1}x_{n+1,l}\right) \, ,  \label{recurr}
\end{equation}
where we introduce the function 
\begin{equation}
y(x)=\frac{e^{(-1+H)/T}+e^{(1-H)/T}x}{e^{(1+H)/T}+e^{(-1-H)/T}x}.
\label{y(x)}
\end{equation}
If a spin $S_{n+1,l}$ 
is on a dead end, then $x_{n+1,l}=1.$ Note that at $H>0$, 
$x_{n,l}\leqslant 1$, while $x_{n,l}\geqslant 1$ for $H<0$. For $H>0$, it is
convenient to introduce 
$x_{n,l}=\exp (-h_{n,l})$. Here, 
at a given $n$, 
$h_{n,l}$ are positive and independent random parameters, which play the
role of random effective fields acting on a spin in the $n$-th shell from
neighboring spins in the $n+1$-th shell. Then, Eqs. (\ref{M}) and (\ref
{recurr}) take the form 

\begin{eqnarray}
M &=&\frac{e^{2H/T}-\exp (-\sum_{l=1}^{k_0}h_{1,l})}{e^{2H/T}+\exp
(-\sum_{l=1}^{k_0}h_{1,l})}\text{,}  \label{M2} \\
\text{ }h_{n,j} &=&-\ln \left\{y\left[\exp
\left(-\sum_{l=1}^{k_{n,j}-1}h_{n+1,l}\right)\right]\right\} \, .
\label{recurr2}
\end{eqnarray}
At dead ends we have $h_{n+1,l}=0.$ At 
$H=0$ in the paramagnetic phase $h_{n,l}=0$, while in the ordered 
phase $h_{n,l}\neq 0.$ Eqs. (\ref{M2}) and (\ref{recurr2}) 
determine the magnetization $M$ of a graph as a function of $T$ and $H$. We
emphasize that these equations are valid for any tree-like graph. 

While deriving the recurrence relations, we started from some spin $S_0$ and
then made the recurrence steps along sub-trees. 
While solving the recurrence relations, 
we start from distant spins, i.e., from large $n$, and descend along
sub-trees to the spin $S_0$. The recurrence steps Eq. (\ref{recurr})
converges exponentially quickly. Therefore, we can set for the starting
spins $h\approx 0$ . In the limit $n\rightarrow \infty $ the parameter $%
h_{1,i}$ is the fixed point of the recurrence steps. The thermodynamic
behavior is determined by this fixed point, which is reached from the
neighborhood of $h=0$.

Let us average the magnetization $M$ over the ensemble of random graphs with
a degree distribution $P(k)$. Introducing the distribution function of 
$h_{n,l}$, $\Psi_n(h)=\langle\delta (h-h_{n,l}\rangle$, and its Laplace
transform $\widetilde{\Psi }_n(s)=\int_0^\infty dh\,e^{-sh}\Psi_n(h)$ we
obtain from Eq. (\ref{M2}) 
the average magnetic moment

\begin{equation}
\langle M\rangle = \sum_kP(k)\!\int _0^\infty \!\!dh \frac{e^{2H/T}-e^{-h}}{%
e^{2H/T}+e^{-h}}\int _{-i\infty }^{i\infty }\frac{ds}{2\pi i} e^{sh} 
\widetilde{\Psi }_1^k(s) \, .  \label{<M>}
\end{equation}
Here $P(k)$ is the probability that the 
vertex $0$ has $k$ 
connections. Equation (\ref{recurr2}) gives the recurrence relation between $%
\widetilde{\Psi }_n(s)$ and $\widetilde{\Psi }_{n+1}(s)$:

\begin{equation}
\widetilde{\Psi }_n(s) = \sum_k\!\frac{P(k)k}{\left\langle k\right\rangle }
\!\!\int _0^\infty \!\!dh\,y^s(e^{-h})\!\!\int _{-i\infty }^{i\infty }\! 
\frac{ds^{\prime }}{2\pi i} e^{s^{\prime }h}\widetilde{\Psi }%
_{n+1}^{k-1}(s^{\prime })  \label{recurr-psi}
\end{equation}
with $\widetilde{\Psi }_{n+1}(s)=1$ at a dead end. Here $P(k)k/\langle
k\rangle$ is the probability that the the neighbor in the $n+1-$th shell of
a vertex from the $n-$th shell has $k$ connections. 
We start from distant spins with $\widetilde{\Psi }_{n+1}(s)\approx 1$ and
large $n$ and make recurrence steps toward smaller $n $ until we reach the
fixed point. In the limit $n\rightarrow \infty $, the recurrence procedure
converges to $\widetilde{\Psi }(s)$. The fixed point $\widetilde{\Psi }(s)$
is a solution of Eq. (\ref{recurr-psi}) in which $\widetilde{\Psi }_n(s)$
and $\widetilde{\Psi }_{n+1}(s)$ are replaced by $\widetilde{\Psi }(s)$.
Then Eq. (\ref{<M>}) with $\widetilde{\Psi }_1(s)\rightarrow \widetilde{\Psi 
}(s)$ gives the exact expression for $\langle M\rangle$. At $H=0$ in the
paramagnetic phase we have $\widetilde{\Psi }(s)=1$.

Let us find the exact critical temperature, $T_c$. Consider a starting
function $\widetilde{\Psi }_n(s)=\exp (-s\delta )$, where $\delta$ is small: 
$0<\delta \ll 1$, $s\delta \ll 1$. After the first $m$ recurrence steps, we
obtain $\widetilde{\Psi }_{n-m}(s)=\exp (-s\delta f^m)$, where $%
f=\left\langle k(k-1)\right\rangle \left\langle k\right\rangle ^{-1}\tanh
(1/T)$. For $f<1$, the recurrence steps lead to the fixed point $\widetilde{%
\Psi }(s)=1$, which corresponds to the paramagnetic phase. $f=1$ at a
certain temperature $T_c$ which satisfies the condition $\left\langle
k(k-1)\right\rangle \left\langle k\right\rangle ^{-1}\tanh (1/T_c)=1$, so
that at $T<T_c$ we have $f>1$ and the recurrence steps lead away from the
point $\widetilde{\Psi }(s)=1$. One sees that $T_c$ is given by Eq. (\ref{Tc}%
). Thus, the estimate (\ref{Tc}) is exact. This result is valid for any
degree distribution with $\langle k^2\rangle<\infty$. 
In this range, the average number of the second nearest neighbors is $z_2 =
\langle k^2\rangle - \langle k \rangle$ \cite{nsw00} ($z_1\equiv\langle k
\rangle$), so

\begin{equation}
\tanh \frac{1}{T_c} = \frac{z_1}{z_2} \, .  \label{plus}
\end{equation}

At zero temperature we find the solution $\widetilde{\Psi }%
(s)=t_s+(1-t_s)\delta _{s,0}$, where $\delta _{s,0}=1$ at $s=0$ and $\delta
_{s,0}=0$ at $s\neq 0$, $t_s$ is the smallest root of the equation $%
t_s=\sum_kP(k)k$ $t_s^{k-1}$. This solution gives $\langle M\rangle = 
1-\sum P(k)t_s^k$, 
which is exactly the size of the giant connected component of the network.
This result is evident. Indeed, 
only vertices in the giant connected component have non-zero spontaneous
moment, which is equal to 1 at $T=0$.

{\em Ansatz}.---To study the critical behavior of thermodynamic quantities,
we develop an ansatz in the spirit of the `effective medium' approach. 
Notice that the right-hand sides of Eqs. (\ref{M2}) and (\ref{recurr2})
depend only on the sum of the independent and equivalent random variables $%
h_{n,j}$. So, let us use the following ansatz

\begin{equation}
\sum_{l=1}^kh_{n,l}\approx kh + {\cal O}(k^{1/2}) \, ,  \label{approx}
\end{equation}
where $h\equiv \langle h_{n,l}\rangle$ is the average value of the
`effective field' acting on a spin. 
The larger $k$ the better this approximation. Therefore, the most
`dangerous' highly connected spins 
are taken into account 
in the best way. 
With this ansatz, 

\begin{equation}
\left\langle M\right\rangle =\sum_kP(k)\frac{e^{2H/T}-e^{-kh}}{%
e^{2H/T}+e^{-kh}} \, .  \label{M-av}
\end{equation}
Applying the 
ansatz (\ref{approx}) to Eq. (\ref{recurr2}) yields

\begin{equation}
h=-\left\langle k\right\rangle ^{-1}\sum_kP(k)k\ln y[e^{-(k-1)h}]\equiv G(h)
\, .  \label{h-av}
\end{equation}
$h$ plays the role of the order parameter. At $H=0$, $h=0$ above $T_c$ and
is non-zero below $T_c$. 

{\em Thermodynamic quantities}.---Let us describe the critical behavior of
the thermodynamic quantities of the Ising model on the infinite networks. 
For this, one must solve the equation of state (\ref{h-av}) near $T_c$ at $%
H=0$. 

{\it The case} $\langle k^4\rangle <\infty $. The expansion of $G(h)$ over
small $h$ has the form: $G(h)=g_1h+g_3h^3+\ldots $. 
For brevity, we do not present exact expressions for the coefficients $g_i$. 
Substituting this expansion into Eq. (\ref{h-av}) determines 
$T_c$ and the order parameter $h\approx a\tau ^{1/2}$ as a function of the
reduced temperature $\tau =1-T/T_c$, where $a=[12\left\langle
k(k-1)\right\rangle ^2/(\left\langle k\right\rangle \left\langle
k(k-1)^3\right\rangle T_c]^{1/2}$. Note that the critical temperature $T_c$
that follows from our ansatz coincides 
with the exact result (\ref{Tc}). At small $h$ and $H=0$ Eq. (\ref{M-av})
gives the spontaneous moment $\left\langle M\right\rangle \approx
\left\langle k\right\rangle a\tau ^{1/2}/2$. The magnetic susceptibility can
be calculated from Eq. (\ref{M-av}), by differentiating over $H$ and then
taking into account the dependence of $h$ on $H$ from Eq. (\ref{h-av}). We
obtain $\chi (H=0)\approx (\left\langle k\right\rangle ^3/(2\left\langle
k^2\right\rangle \left\langle k(k-2)\right\rangle ))\tau ^{-1}$. At $T>T_c$
the susceptibility has the same behavior $\chi (0)\sim (T/T_c-1)^{-1}$ but
with the double prefactor.

In our ansatz, the average internal energy 
per spin, $\langle E\rangle =\left\langle -J\sum_{\left\langle
ij\right\rangle }S_iS_j\right\rangle _T/N$ (the average is thermodynamic and
over the ensemble of graphs), at $H=0$ 
is 

\begin{eqnarray}
\left\langle E\right\rangle &=&-\frac 12\left\langle k\right\rangle \coth
(2/T)+  \nonumber \\
&&+\frac 1{2\sinh (2/T)}\sum_kP(k)k\frac{e^{-h}+e^{-(k-1)h}}{1+e^{-kh}} \, .
\label{E}
\end{eqnarray}
Substituting here $h\approx a\tau ^{1/2}$, we find that the specific heat $%
C=d\left\langle E\right\rangle /dT$ has a jump at $T=T_c$:

\begin{equation}
\Delta C=\left\langle k(k-2)\right\rangle \left\langle k^2\right\rangle
a^2/(8\left\langle k\right\rangle ) \, .  \label{C1}
\end{equation}
This jump disappears as $\langle k^4\rangle$ approaches $\infty$. When $%
\langle k^4\rangle$ diverges, we consider power-law degree distributions, $%
P(k) \propto k^{-\gamma}$.

{\it The case }$\gamma=5$. 
The divergence of $\left\langle k^4\right\rangle$ leads to a logarithmic
singularity of the function $G(h)$: $G(h)=g_1h+g_3h^3\ln (1/h)+\ldots$.
Solving Eq. (\ref{h-av}) yields

\begin{equation}
h,M\sim \tau ^{1/2}/(\ln \tau ^{-1})^{1/2},\text{ }\delta C\sim 1/\ln \tau
^{-1},\text{ }\chi \sim \tau ^{-1} \,  \label{g5}
\end{equation}
[for any $\gamma$, we have $\delta C(T>T_c)=0$]. 
Note that the critical behavior of $\chi$ is not changed.

{\it The case }$3<\gamma <5$. The function $G(h)$ has two leading terms: $%
G(h)\approx g_1h+g_3h^{\gamma -2}$. Solving the equation of state (\ref{h-av}%
) yields

\begin{equation}
h,M\sim \tau ^{1/(\gamma -3)},\text{ }\delta C\sim \tau ^{(5-\gamma
)/(\gamma -3)},\text{ }\chi \sim \tau ^{-1}.  \label{g3-5}
\end{equation}
Notice that for $\gamma<5$ 
critical exponents of the magnetization and specific heat differ from the
standard ones. 
For $\gamma >4$, exponent $\alpha$ of the specific heat $(\delta C\sim \tau
^{-\alpha })$ is above $-1$. Hence the second derivative of the free energy
over $T$ diverges at $T_c$. Therefore, the phase transition is of the second
order in Ehrenfest's classification. For $\gamma <4$, we have $\alpha
=-(5-\gamma )/(\gamma -3)<-1$ and the transition turns to be of a higher
order. The order of the transition tends to infinity as $\gamma \rightarrow
3 $. Nevertheless, for $\gamma>3$, the susceptibility obeys the Curie law.

{\it The case }$\gamma =3$. 
Here, $\left\langle k^2\right\rangle $ 
diverges. 
Formally speaking, this leads to the infinite critical temperature for the
infinite networks [see Eq. (\ref{Tc})]. In any finite network, $\left\langle
k^2\right\rangle <\infty $, and the critical temperature is finite, although
it may be very high, $T_c\cong \langle k^2\rangle /\langle k\rangle $ (see
below). 
We consider temperatures, which are much less than this critical
temperature, but where $h\ll 1$, so $T\gg 1$. 
Using, for brevity, the continuum approximation for the degree distribution, 
we obtain $G(h)\approx (\left\langle k\right\rangle h/2T)\ln [2/(\langle
k\rangle h)]$. Then, 

\begin{eqnarray}
h &\approx &(2/\langle k\rangle )e^{-2T/\left\langle k\right\rangle },\text{
\quad }M\approx e^{-2T/\left\langle k\right\rangle },\text{ \quad } 
\nonumber \\
\delta C &\sim &T^2e^{-4T/\left\langle k\right\rangle },\text{ \quad }\chi
\sim T^{-1}.  \label{gamma3}
\end{eqnarray}
Without the continuum approximation, we have, instead of $\langle k\rangle $
in the exponentials, a constant that is determined by 
the complete 
$P(k)$. Equation (\ref{gamma3}) describes the behavior of thermodynamic
quantities at modest temperatures in the situation, where the phase
transition is of infinite order and at `infinite temperature'. Notice the
paramagnetic dependence $\chi \propto 1/T$. 
Note that the temperature dependence $M\propto \exp [-2T/\langle k\rangle ]$
coincides with the result of the simulation \cite{ahs01} for the
Barab\'{a}si-Albert model ($\gamma =3$) [see Fig. 1 (a) from this paper].

{\it The case} $2<\gamma <3$. Again $T_c$ for large networks is very high. 
Using the expansion $G(h)\approx gh^{\gamma -2}/T$, we find, in the range $%
1\ll T\ll T_c$, the following behavior

\begin{equation}
h,\!M\sim T^{-1/(3-\gamma )}\!,\ \delta C\sim T^{-(\gamma -1)/(3-\gamma
)}\!,\,\chi \sim T^{-1}  \label{gamma2-3}
\end{equation}
[compare with Eq. (\ref{gamma3})]. 
Notice that the susceptibility keeps the paramagnetic temperature dependence.

{\em Finite-size effect}.---Equation (\ref{Tc}) shows that $T_c$ diverges
when $\langle k^2\rangle \to \infty$. 
In finite networks, $\langle k^2\rangle $ is finite because of the
finite-size cutoff of the degree distribution. In scale-free networks, it is
estimated as $k_{cut}\sim k_0N^{1/(\gamma -1)}$, where $N$ is the total
number of vertices in a network, $k_0$ is a `minimal degree' or the lower
boundary of the power-law dependence, and $\langle k\rangle \approx
k_0(\gamma -1)/(\gamma -2)$. Then, actually, repeating estimations from Ref. 
\cite{dm01c} (see more detailed discussions in Refs. \cite{cbh02,pv02}), we
obtain

\begin{eqnarray}
&& T_c\approx \frac{\langle k\rangle \ln N}4\ \ \ \ \ \ \ \ \ \ \ \ \ \ \ \
\ \ \ \ \ \ \ \ \ \ \ \ \ \ \ \mbox{at}\ \gamma =3 \,,  \nonumber \\
&& T_c\approx \frac{(\gamma -2)^2}{(3-\gamma )(\gamma -1)}\langle k\rangle
N^{(3-\gamma )/(\gamma -1)}\ \ \mbox{for}\ 2<\gamma <3 .  \label{2}
\end{eqnarray}
The first expression can be compared with the simulation \cite{ahs01} of the
Ising model on the Barab\'{a}si-Albert growing network ($\gamma =3$) with
minimal degree $k_0=5$, so $\langle k\rangle =10$. From Eq. (\ref{2}) $%
T_c\approx 2.5\ln N$ follows. The simulation \cite{ahs01} yields $T_c\approx
2.6\ln N-3$. Recall that our results were obtained for the completely
uncorrelated network, and correlations in growing networks are extremely
strong. 

{\em Discussion}.---Our results may be compared with those for percolation
on such networks \cite{cbh02} and the disease spread within them \cite{pv01}%
. Of course, the problems are distinct but one finds a great resemblance
(the super-stability of long-range order when $\langle k^2\rangle \to \infty$%
, phase transitions of higher order, etc.). All these anomalous features are
determined by fat tails in the distributions of connections. Our final
results were presented for uncorrelated networks. However, the quantitative
agreement with the simulation \cite{ahs01} of correlated nets shows that
they are applicable in much more general situations. Furthermore, our
analytical results can be easily generalized.

S.N.D. thanks PRAXIS XXI (Portugal) for a research grant PRAXIS
XXI/BCC/16418/98. S.N.D and J.F.F.M. were partially supported by the project
POCTI/99/FIS/33141. A.G. acknowledges the support of the NATO program
OUTREACH. We also thank A.N. Samukhin and D. Stauffer for useful
discussions.  \newline

\noindent
$^{*}$ E-mail address: sdorogov@fc.up.pt \newline
$^{\dagger }$ E-mail address: goltsev@pop.ioffe.rssi.ru \newline
$^{\ddagger }$ E-mail address: jfmendes@fc.up.pt

\end{multicols} 

\end{document}